\documentstyle[epsf,prl,twocolumn,aps]{revtex}

\newcommand{\be}{\begin{equation}}
\newcommand{\ee}{\end{equation}}
\newcommand{\bea}{\begin{eqnarray}}
\newcommand{\eea}{\end{eqnarray}}

\newcommand{\la}{\langle}
\newcommand{\ra}{\rangle}

\newcommand{\p}{\partial}

\begin {document}
\bibliographystyle {plain}
\title{Logarithmic Operators in the Theory of Plateau Transition}
\author{Ian I. Kogan  and  Alexei M. Tsvelik}
\address{Department of Physics, University of Oxford, 1 Keble Road, Oxford, 
OX1 3NP, UK}
\date{\today }
\maketitle

\begin{abstract}
\par
We  show that $SL(2;C)/SU(2)$ model which had been recently 
proposed  to  describe  the behaviour of the local densities of
states at the plateau transition in  Integer Quantum Hall effect,
has  logarithmic operators. They unusual properties are studied in
this letter.

\end{abstract}
OUTP-99-08S 
PACS numbers: 75.10 Jm, 75.40 Gb.  
\sloppy


\par
 In our previous paper \cite{Five99} we have discussed a problem of
plateau transition in Integer Quantum Hall effect. We have provided evidence 
to suggest that the correlation functions of
the local densities of states (DOS) at the critical point 
 may be obtained from the {\em critical} $ SL(2;C)/SU(2)$
Wess-Zumino-Novikov-Witten (WZNW) model, 
in which the level $k' = - (k + 2)$ ($k > 0$) is fixed by the known scaling 
dimensions.  The action of this model is given by 
\begin{equation}
\label{slaction}
S_0=\frac{(k+2)}{4\pi}\int d^2x \left[4\partial
\theta{\bar\partial}\theta+e^{2\theta}\partial\mu{\bar\partial}\mu^{\ast}\right]
\label{sltwo}
\end{equation}
This model has numerous applications in field theory 
 including the theory  of disorder (see, for example   
\cite{Caux:disfer} and references therein).
Among unconventional features displayed by this model is the presence
 of non-trivial primary fields  with zero scaling dimension. These
 operators are likely to play very important role in the applications.
 Thus, in \cite{Five99} we have suggested that to study    correlation
functions of local DOS one has to  consider the effective action 
\begin{equation}
S=S_0+\eta\int d^2x\, \Psi_0({\bf x}),
\end{equation} 
where $\Psi_0$ is a non-trivial primary field of scaling dimension
zero. 
We have argued that in order to obtain the consistent results 
 for  
 the correlation functions at the critical point, one has to execute 
the $\eta \rightarrow 0$ limit. This  includes 
expansion in powers of $\eta$ keeping 
only the first non-vanishing term.
In this procedure  the two-point correlation function of the $q^{th}$
powers of the local DOS  is  expressed  as the 
 limit of the  {\em four-point} correlation function:  
\bea 
\overline{\rho^q({\bf r}_1)\rho^q({\bf r}_2)}= \\
\lim_{|{\bf r_3}|\rightarrow 0, |{\bf r}_4|\rightarrow\infty}\langle\rho^q({\bf
r}_1)\rho^q({\bf r}_2)\Psi_0({\bf r}_3)\Psi_0({\bf r}_4)\rangle
\nonumber
\eea
where we use the overbar to denote disorder averaging, and use the
angular brakets to denote correlation functions of our field
theory. Suggesting this  procedure we have assumed that $\Psi_0$ was a
 relevant operator such that at any finite $\eta$ its presence in the
 effective action generates the  scale $L_{\eta} \sim \eta^{-1/2}$. 

 In
 this paper we show that the theory  possesses another field with zero
 scaling dimension which presence in the effective action does not
 violate the conformal invariance. A similar example has already been
 considered in   \cite{Five99} (see Appendix 6) where it was shown
 that  at  $k=2$ kinetic the energy perturbation  model does not
 violate conformal invariance of model (\ref{sltwo})0). 
The corresponding   operator
has the Operator Product Expansion (OPE)   $ \Psi_0 (z) \Psi_0(0) =
 \Psi_0(0)$, and hence   has  zero
  norm. Such OPE  is essential for preservation of  the conformal
 invariance of the theory.

 There is a degeneracy in the present problem: besides the 
nontrivial operator with dimension zero there is always a unity
operator.  One can think that the pair of degenerate  operators  form 
the logarithmic pair \cite{Gurarie}. In such a pair one of the operators 
always has zero norm \cite{CKT}:
\begin{eqnarray}
\langle C(x) D(y)\rangle &&= 
\langle C(y) D(x) \rangle  = \frac{c}{(x-y)^{2\Delta_C }}\nonumber \\
\langle D(x) D(y)\rangle &&= 
 \frac{1}{(x-y)^{2\Delta_C}} \left(-2c\ln(x-y) + d\right)
\nonumber \\ 
\langle C(x) C(y)\rangle  &&= 0
\label{twopointcorrfunction}
\end{eqnarray}
where in our case $\Delta = 0$
   
The purpose of this letter  is to investigate the properties of  
zero-dimensionmal operators  $SL(2,R)$ model and to see if there are
 logarithmic operators in the theory. 
Let us note that logarithmic  operators in 
 $SL(2,R)$  model have been discussed before (see for example \cite{BKM})
 but  not particularly   zero dimensional ones.  

Primary fields of model (\ref{slaction}) are tensors from the
representations of the $SL(2,R)$ group. Each representation is
labeled by its angular momentum $j$. The conformal weights
are 
\begin{equation}
\label{conf}
h_{j}=-j(j+1)/k
\end{equation}
 The multi-point correlation function of primary fields belonging to
 representations with angular momentum $j_a$ is given by the following
 Knizhnik-Zamolodchikov (KZ) equation \cite{KLS},\cite{FZ}:
\bea
\left\{ k\frac{\p}{\p z_i} - \sum_{j \neq
 i}^N\frac{L_{ij}}{z_{ij}}\right\}
G(1\cdots N) = 0 ~~~~~ 
\label{KZequation}
\\
 L_{ij} = (y_{ij})^2
\frac{\p^2}{\p y_i\p y_j} +  
2y_{ij}\left(j_j\frac{\p}{\p y_i} - j_i\frac{\p}{\p y_j}\right) -
 2j_ij_j
 \nonumber 
\eea
Here the parameter $y$ parametrizes the $SL(2,R)$ group variables and one gets 
the following representation for the $SL(2,R)$ generators
\bea
t^+ = \frac{\partial}{\partial y}, ~~~
t^3 = y\frac{\partial}{\partial y}-j,~~~ 
t^+ = y^2\frac{\partial}{\partial y}- 2j y
\eea
In this parametrization 
 a matrix  operator $\Phi_{k,\bar k}$ is mapped on a function of
 complex variable $\Phi(y,\bar y)$ (we omit here the dependence on spatial coordiantes
 $z $ and $\bar{z}$) as
\be
\Phi^{(j), (\bar{j})}(y, \bar{y}) = \sum_{k,\bar k }y^{j + k}
{\bar y}^{\bar j + \bar k}\Phi^{(j), (\bar{j})}_{k,\bar k}
\label{Fofy}
\ee

 Let us discuss  now  the case when all operators have $j = 0$. 
 The  equation  (\ref{KZequation}) for  $j = 0$ becomes 
\bea
\left\{ k\frac{\p}{\p z_i} -  
\sum_{j \neq
 i}^N\frac{1}{z_{ij}}\left[(y_{ij})^2\frac{\p^2}{\p y_i\p
 y_j}\right]\right\}G(1\cdots N) = 0 \label{KZ0}
\eea
 We shall concentrate on holomorphic dependence on $z$ and $y$ only,
 the restoration of $\bar{z}$ and $\bar{y}$ dependence is 
straightforward.
 Let us study first of all the case of $N=2$. The
 two-point functions $G_2(z,y) = <\Psi_{0} (z,y)
\Psi_{0}(0,0)>$ which obeys the equation
\bea
\left\{ k\frac{\p}{\p z} + \frac{y^2}{z}  
\frac{\p^2}{\p y^2} \right\}G_2(z,y) = 0 \label{KZ02point}
\eea
 Using the Ward identities 
\bea
\sum_{n=1}^N \left( y_n^{l+1}\frac{\p}{\p y_n}
 - (l+1) j_n y_n^l \right)<\Phi^{j_1} (1) ...
\Phi^{j_n} (n)> = 0, \nonumber \\
l = -1,0,+1 ~~~ (1) = (z_1, y_1),..., (n) = (z_n,y_n)
\eea
one  gets for $j_n =0$  very simple constraints
\bea
\sum_{n=1}^N y_n^{l+1}\frac{\p}{\p y_n} G(1...N) = 0, ~~
l = -1,0,+1 ~~~
\label{ward}
\eea
and from $l=0$  constraint  we see that $y \frac{\p}{\p y}G_2(z,y) = 0$ 
which means that $\frac{\p}{\p z} G_2(z,y) = 0$ and the two-point function is 
a constant.

For $N = 4$  the Ward identities  (\ref{ward}) and similar Virasoro
$L_{\pm 1}, L_{0}$ constraints  preserve the invariance of correlation
functions under projective transformations in $z$ and $y$
planes. Hence the four point function
$G(1,...4) $ depends only on the anharmonic ratios $z$ and $t$ where
\[
z = \frac{z_{32}z_{41}}{z_{31}z_{42}},
\quad t = \frac{y_{32}y_{41}}{y_{31}y_{42}}
\] 
 and Eq.(\ref{KZ0})  for $G(1,...4)= {\cal F}(z,t)$ 
 is simplified by  the projective nvariance 
\be
- \frac{t^2}{z}\p_t[(t - 1)\p_{t}{\cal F}] + \frac{(t - 1)^2}{z
- 1} \p_t(t\p_t{\cal F}) + 
 k\p_z{\cal F} = 0 \label{Eq}
\ee
Let us note that this equation is invariant under transformation
$ z \rightarrow 1-z, ~ t \rightarrow 1-t$, which means that for any solution
 ${\cal F}(z,t)$ there is another one ${\cal F}(1-z,1-t)$.

One  particularly simple solution for the holomorphic part is given by 
\bea
G(1,...4) = A\ln(z/t^k) + B\ln[(z - 1)/(t - 1)^k] \label{4point}
\eea

We see that we have logarithmic singularities in this correlation function.
 Moreover,  besides  logarithms 
 of $z$ we also have logarithms  of $t$. The former ones 
 are due to the Jordan cell structure  of the Virasoro algebra \cite{Gurarie}
 whereas  the latter  ones  are  due to the Jordan cell structure of the
 Kac-Moody algebra \cite{KLS}, \cite{KL}. Hence  expansion (\ref{Fofy})
 must be modified 
 \be
\Phi^{(j)}(y,z) = \sum_{k } y^{j+k} \left[
\Phi^{(j)}_{k,}(z) + \tilde\Phi^{(j)}_{k}(z) \ln y \right] 
\ee

One can choose constants $A$ and $B$ to get two types of invariant
solutions: for $B = 0$  we get the  solution which changes sign  under 
permutation $z,t \rightarrow 1/z.1/t$ which corresponds to the
permutation $1
\rightarrow  2$ or $3 \rightarrow 4$:
\bea
\la F_1(1)F_1(2)F_2(3)F_2(4)\ra = \ln\left|z/t^k\right| \label{fermions}
\eea
Choosing $A = - 2B$ we get the solution with is invariant under 
$z,t \rightarrow z^{-1}, t^{-1}$: 
\bea
\la B_1(1)B_1(2)B_2(3)B_2(4)\ra = \nonumber \\
\ln\left[\frac{|z|}{|1 - z|^2}\right] -
\ln\left[\frac{|t|^k}{|1 - t|^{2k}}\right] \label{bosons}
\eea
It is easy to check that there are no solutions invariant under
permutations of all operators. Therefore we are lead to the conclusion
that there are two  types of fermionic fields $F_a$  and two types of
bosonic fields $B_a$ ( $a = 1,2$) such that their 
4-point 
correlation functions are given by Eqs.(\ref{fermions},\ref{bosons}). 
Hence there is  a hidden $SU(2)$ symmetry in this problem.

 From (\ref{fermions}) we derive the following Operator Product Expansion
(OPE) for 
$F_1, F_2$ :
\bea
F_a (z + \epsilon_z, y + \epsilon_y)F_b(z,y) = \epsilon_{ab}<F F>
\nonumber\\
+  \epsilon_{ab}\left[\ln|\epsilon_z|
 C_z(z,y) + \ln|\epsilon_y|
 C_y(z,y) + D(z,y) + ...\right] 
 \label{FFope}
 \eea
where the  $\epsilon_{ab}$ structures are dictated by $SU(2)$ 
symmetry and dots stand for terms vanishing at $\epsilon \rightarrow 0$. 
From this OPE one can see that the fusion of $F_1$ and $F_2$  generates a 
logarithmic triad $C_z, C_y, D$:
 \bea
\la C_a C_b\ra =
 0,  \nonumber\\ 
\la  C_z(1)D(2)\ra = 1, \quad \la C_y(1)D(2)\ra = - k, \\
\la  D(y,z)D(0,0)\ra = 2\ln|y^k/z| \nonumber
\eea

Let us note that this triad  actually represents two pairs: the 
Virasoro logarithmic pair
 \cite{Gurarie}  $C_z, D$ which is mixed under the conformal transformation 
 $z \rightarrow \lambda z$ as
 \bea
 D \rightarrow D - \ln\lambda C_z, ~~~ C_z   \rightarrow C_z
 \eea
and the Kac-Moody logarithmic pair \cite{KLS} \cite{KL}  $C_y, D$ which is mixed
 under the SL(2,R) transformation $y \rightarrow \epsilon y$ generated by $t^3 =
 y(\partial/\partial y)$:
 \bea
 D \rightarrow D - \ln\epsilon C_y, ~~~ C_y   \rightarrow C_y
 \eea

For the bosonic operators one  gets the following OPE from 4-point
function (\ref{bosons}):
\bea
&B_a&(z + \epsilon_z, y + \epsilon_y)B_b(z,y) = (2\delta_{ab} - 1)\nonumber\\
&\times&\left[\ln|\epsilon_z|
 C_z(z,y) + \ln|\epsilon_y|
 C_y(z,y) + D(z,y) + ...\right]
\eea

In this paper we would like to concentrate on the fermionic sector only. 
 The role of   bosonic fields $B_a$  and they  relation  to the
 operators $\Psi_0$ used in \cite{Five99} will be discussed elsewhere.

It is logical to suggest that $\la C_a\ra = 0$. Therefore the
two-point correlation function of $F_a$ is constant which is
consistent with the result derived from Eq.(\ref{KZ02point}). To get a
non-trivial coordinate dependence  one has to define the new two-point
function putting in it  a vacuum insertion of
$D$:
\bea
\la F_1(1)F_2(2)D(\infty)\ra = \ln|z_{12}/y^k_{12}|
\eea

We also can study the 6-point functions, in which case the answer 
can be presented  as an antisymmetrized product of 2-point functions
and 4-point functions :
\bea
~~\la F_1(1)F_1(2)F_1(3)F_2(4)F_2(5)F_2(6)\ra =  \nonumber \\
+\la F_1(1)F_2(6)\ra\la F_1(2)F_1(3)F_2(4)F_2(5)\ra \nonumber \\
-\la F_1(1)F_2(5)\ra\la F_1(2)F_1(3)F_2(4)F_2(6)\ra \nonumber \\
-\la F_1(1)F_2(4)\ra\la F_1(2)F_1(3)F_2(6)F_2(5)\ra \nonumber \\
-\la F_1(2)F_2(6)\ra\la F_1(1)F_1(3)F_2(4)F_2(5)\ra \label{6fermions} \\
+\la F_1(2)F_2(5)\ra\la F_1(1)F_1(3)F_2(4)F_2(6)\ra \nonumber \\
+\la F_1(2)F_2(4)\ra\la F_1(1)F_1(3)F_2(6)F_2(5)\ra \nonumber \\
-\la F_1(3)F_2(6)\ra\la F_1(2)F_1(1)F_2(4)F_2(5)\ra \nonumber \\
+\la F_1(3)F_2(5)\ra\la F_1(2)F_1(1)F_2(4)F_2(6)\ra \nonumber \\
+\la F_1(3)F_2(4)\ra\la F_1(2)F_1(1)F_2(6)F_2(5)\ra \nonumber
\eea
Using the fact that the two-point functions $\la F_1(i)F_2(j)\ra$
 is a constant (it may be  even zero, but we do not know this  with 
certainty and this is not important) and the four-point function is given by 
(\ref{fermions})
\bea
\la F_1(a)F_1(b)F_2(c)F_2(d)\ra = \ln \left|\frac{z_{bc} z_{ad}}
{z_{ac} z_{bd}} \right| - k \ln \left|\frac{y_{bc} y_{ad}}
{y_{ac} y_{bd}}\right|
\eea
we can   easily  see  that the six-point function is identically zero. 
The proof of this is the following. 
We can see that each pair of indices $a$ and $b$ where the first one is $1,2,3$ 
and the second one is $4,5,6$ appears precisely in four terms in (\ref{6fermions}). 
For example the pair $3,4$  appears in a first, second, fourth and fifth four-point 
correlation functions, which means that the terms $\ln |z_{34}|$ and 
$\ln |y_{34}|$ will enter four times. A simple inspection of (\ref{6fermions}) shows 
that two times they enter with sign plus and two times with sign minus and the total 
contribution is zero.  In the same way we can check each pair of indices, for example 
the pair $1,5$ enters fourth, six, seventh and ninth correlation functions and again 
the total contribution is zero, etc.

Thus we see that the the six-point correlation function is zero. 
 It is not difficult to generalize these  solutions for general $N =
4M$ and $N = 4M + 2$. In the  first case one has to decompose $N$ points 
in groups of four in all possible ways and take  a  symmetrized (bosons) 
or antisymmetrized (fermions) sum of products  of solutions (\ref{4point}). 
In the second case we  multiply this by two-point correlation 
functions $<F_1 F_2>$,  but since  they are constants it is essentially the 
same procedure. One can show that  for fermions 
all these correlation functions
are identically  zero. 
We have checked it explicitly for 8-point functions.  The structure 
of antisymmetrization in a higher dimensional case is such that 
each pair appears 
in an even number of  four-dimensional correlation functions 
half of which have negative 
signs.

 It is  most remarkable that all higher-order  fermionic 
correlation functions vanish. Since the $C,D$-operators of the triad
can be generated via fusions of the fermions (\ref{FFope}), this result means that
all higher-order correlation functions  of the logarithmic fields vanish.
Not only three-point functions are trivial
 \bea
\la C_a~C_b~C_c\ra = 0, \quad \la C_a~C_b~ D\ra = 0, \nonumber \\
 \quad \la C_a~D~D\ra = 0,
\la D~D~D\ra = 0
\eea
where indices $a,b,c,$ for zero norm operators $C$ means $y$ or $z$, but also all 
four-point functions are
\bea
\la CCCC\ra = 0, \quad \la CCCD\ra = 0, \quad \la CCDD\ra = 0, \nonumber \\
\la CDDD\ra = 0,
\quad \la DDDD\ra = 0, 
\eea
 
From here we can dedudce that OPE of $C$ and $D$ may have only the $C$
operator (besides unity operator $I$):
\bea
C_a ~ C_b &=& F_{abc}C_c \nonumber \\
C_a ~ D   &=& <C_a D> I + F_{ab} C_b \\
D ~ D  &=& <D D> I + F_{a} C_a \nonumber
\eea

 This is 
precisely what one  needs to be able to 
truncate an expansion in   superpositions of
 $C$ and $D$ operators. Adding a  new $C$ or $D$ term inside a 
 correlation  function does   not change its   scaling behaviour; 
 only the  first two $C$ and $D$ operators matter, not the next ones.
 Thus we conclude that model (\ref{sltwo}) remains conformally 
invariant when one adds to the action the term 
\bea
\eta\int d^2x :F_1F_2:
\eea

 It is quite a challenge to fully understand the physical
 meaning of the operators discussed in this paper. The reason is that
 our knowledge of relationship
 between observables in the Quantum Hall problem and fields of model
 (\ref{sltwo}) is very limited.  The only thing we can be sure about
 is that such operators are important. This certainty is supported by
 results obtained for  other 
disordered systems  (see   
\cite{GL99},\cite{cardy99}), where operators with zero scaling  dimensions
also  play very important role. Hopefully, better understanding of
 mathematical structure of these theories will help to 
 get an insight into  their physical meaning. It is also possible that
 these operators will play an important role in other physical
 problems where  $SL(2.R)$ WZNW model is important, for example, for 
 gravitational dressing of exactly integrable models and logarithmic
 operators in string theory \cite{BKM}.


\begin{thebibliography}{99}


\bibitem{Five99} M. J. Bhaseen, I. I. Kogan, O. A. Soloviev,
N. Taniguchi and A. M. Tsvelik, cond-mat-9912060.
\bibitem{Caux:disfer}J.-S. Caux, N.~Taniguchi, and A.~M. Tsvelik, 
\newblock Phys. Rev. Lett. {\bf 80}, 1276 (1998); 
Nucl. Phys. {\bf B525}, 621 (1998). J.-S. Caux, Phys. Rev. Lett. {\bf
81}, 4196 (1998).

\bibitem{Gurarie} V. Gurarie, Nucl. Phys. B{\bf 410}, 535 (1993).
\bibitem{CKT} J.-S. Caux, I. I. Kogan and A. M. Tsvelik,
Nucl. Phys. B{\bf 466}, 444 (1996). 
\bibitem{BKM} A. Bilal and I.~I. Kogan, hep-th/9407151,
 Princeton University preprint PUPT-1482 (unpublished);
 Nucl.Phys. {\bf B 449}, 569 (1995);
I.~I. Kogan and N.Mavromatos, Phys. Lett. {\bf B375}
 (1996), 111
\bibitem{KLS}I.~I. Kogan, A.~Lewis, and O.~A. Soloviev,
\newblock Int. J. Mod. Phys. {\bf A13}, 1345 (1998)
\bibitem{KL}I.~I. Kogan and  A.~Lewis, Nucl. Phys. B{\bf 509}, 687
(1998).
\bibitem{FZ} V. Fateev and A. Zamolodchikov, Sov. Nucl. Phys. {\bf
43}, 1031 (1986). 
\bibitem{GL99} V. Gurarie and A. W. W. Ludwig, cond-mat/9911392.
\bibitem{cardy99} J. L. Cardy, cond-mat/9911457.

\end{thebibliography}
\end{document}